\documentclass[
               twocolumn,
               noshowpacs,          
               nopreprintnumbers,     
               aps,                 
               prd,                 
               a4paper,             
               superscriptaddress,  
               nofootinbib,         
               tightenlines,        
               floats,floatfix      
               ]{revtex4-2}

\usepackage{amsmath, amsfonts, amsthm, amssymb, graphicx}
\usepackage{epstopdf}
 \usepackage{slashed}
 \usepackage{graphicx,epsfig}
\usepackage{amsmath}
\usepackage {amssymb}
\usepackage[utf8]{inputenc}
\usepackage{hyperref}
\usepackage{xcolor}
\usepackage[T1]{fontenc}

\newcommand{\be}{\begin{eqnarray}}
\newcommand{\ee}{\end{eqnarray}}
\newcommand{\bea}{\begin{eqnarray}}

\newcommand{\eea}{\end{eqnarray}}

\def\d{\partial}

\begin{document}

\title{Integrable Systems and Spacetime Dynamics}

\author{Marcela C\'ardenas}
\email{marcela.cardenas-at-usach.cl}
\affiliation{Departamento de F\'{i}sica, Universidad de Santiago de Chile, Avenida Ecuador 3493, Estaci\'on Central, 9170124, Santiago, Chile.}

\author{Francisco Correa}
\email{francisco.correa-at-uach.cl}
\affiliation{Instituto de Ciencias F\'isicas y Matem\'aticas, Universidad Austral de Chile, Casilla 567, Valdivia, Chile.}

\author{Kristiansen Lara}
\email{kristiansen.lara-at-usach.cl}
\affiliation{Departamento de F\'{i}sica, Universidad de Santiago de Chile, Avenida Ecuador 3493, Estaci\'on Central, 9170124, Santiago, Chile.}

\author{Miguel Pino}
\email{miguel.pino.r-at-usach.cl}
\affiliation{Departamento de F\'{i}sica, Universidad de Santiago de Chile, Avenida Ecuador 3493, Estaci\'on Central, 9170124, Santiago, Chile.}
\affiliation{Center for Interdisciplinary Research in Astrophysics and Space Exploration (CIRAS), Universidad de Santiago de Chile, Avenida Libertador Bernardo O'Higgins 3363, Estación Central, Chile.}


\date{\today}

\begin{abstract}
It is shown that the Ablowitz-Kaup-Newell-Segur (AKNS) integrable hierarchy can be obtained as the dynamical equations of three-dimensional general relativity with a negative cosmological constant. This geometrization of the AKNS system is possible through the construction of novel boundary conditions for the gravitational field. These are invariant under an asymptotic symmetry group characterized by an infinite set of AKNS commuting conserved charges. Gravitational configurations are studied by means of $SL(2,\mathbb{R})$ conjugacy classes. Conical singularities  and black hole solutions are included in the boundary conditions.

\end{abstract}

\maketitle

\textit{Introduction.} 
Over the years, the significance of integrable systems has been successfully demonstrated and tested in almost all areas of physics. The applications range from optical wave propagation in nonlinear media \cite{fiber} to  nonlocal phenomena \cite{nonlocal}, self-interacting fermion systems \cite{self}, and many other examples. Naturally, the notions of integrability have reached curved space-time theories and general relativity, for instance, by the Ernst equation \cite{ernst} and the inverse scattering method, both being used to solve the Einstein equation in different scenarios \cite{gravsol}. More recently, integrability aspects have been studied in the context of AdS/CFT \cite{adscft}.  In the same line, twistor and self-dual Yang Mills theories find also interesting connections with integrable models \cite{selfdualbianchi, mw}.  The Ward conjecture explores the possibility that all integrable systems find a common origin as reductions of self-dual Yang Mills equations \cite{ward,sfyg} and has been checked for well-known examples such as  Korteweg-de Vries,  sine-Gordon,  nonlinear Schr\"odinger, and the Ablowitz-Kaup-Newell-Segur (AKNS) hierarchy \cite{Ablowitz:1973zz}. The AKNS hierarchy is a fundamental integrable model which encompasses and generalizes all mentioned 1+1 integrable systems and find its own physical applications in different scenarios \cite{book}. In this Letter, we will describe a new relationship between the huge family of integrable models given by the AKNS hierarchy and the spacetime dynamics of general relativity with negative cosmological constant. A broad class of compatible boundary conditions for the gravitational field is constructed on account of the diverse properties of the AKNS system. This  allows us to create a dictionary among gravity and integrable features. Under this spirit, the paper is organized as follows. First we demonstrate how the nonlinear integrable AKNS hierarchy arises as the field equations of general relativity in three dimensions with negative cosmological constant. Then, we show how the conserved quantities inherent of AKNS give rise to a family of nonequivalent boundary conditions for the gravitational field. We finish by showing how the generic solutions scheme of the AKNS system can be mapped into novel gravitational solutions such as black holes. \newline

\textit{Geometrization of AKNS.} 
Let us first consider a three-dimensional spacetime manifold $\mathcal{M}$, foliated by the coordinates $\left(t,\rho,\varphi\right)$, where $t$ represents time and  $\left(\rho,\varphi \right) $ are polar coordinates. The units are chosen such that $t$ and $\rho$ have dimensions of length, whereas the angle $\varphi$ is dimensionless. Assume that the manifold is endowed with a metric field whose line element, given in ADM decomposition $ds^2=-N^2 dt^2 + (N^i dt +dx^i)(N^j dt + dx^j)g_{ij}$, is constructed in the following fashion. The lapse function has the form 
\begin{equation}\label{lapse}
  N^2=\frac{\rho^2}{4\ell^2}\frac{\left(    \Omega^+   \omega^-  + \Omega^- \omega^+   \right)^2}{   \omega^- \omega^+  },
\end{equation}
while the shift vector components are given by
\begin{subequations}\label{shift}
\begin{align}
  N^\rho&=\frac{\rho}{\ell} \left[A^- - A^+  + \frac{1}{2} \left(\xi^+ + \xi^-        \right)    \left(\frac{\Omega^-}{\omega^-}-\frac{\Omega^+}{\omega^+}\right)\right],\\
 N^\phi&=\frac{1}{2\ell}  \left(\frac{\Omega^-}{\omega^-}-\frac{\Omega^+}{\omega^+}\right).
\end{align}
\end{subequations}
Additionally, the spatial metric reads
\begin{equation}\label{gij}
  g_{ij}= \left(
    \begin{matrix}
      \frac{\ell^2}{\rho^2}& - \frac{\ell^2}{\rho}\left(\xi^++\xi^- \right)\\
       - \frac{\ell^2}{\rho}\left(\xi^++\xi^- \right)& \ell^2 \left(\xi^++\xi^- \right)^2+\rho^2 \omega^-\omega^+
    \end{matrix}
\right).
\end{equation}
 The auxiliary functions $\Omega^\pm$ and $\omega^\pm$ are defined as
\begin{equation}
  \Omega^\pm\equiv B^\pm - \frac{\ell^2}{\rho^2}C^\mp,\quad   \omega^\pm\equiv p^\pm + \frac{\ell^2}{\rho^2}r^\mp,
\end{equation}
where $\ell$ stands for the AdS$_3$ radius. The spacetime metric components contain two sets of dimensionless functions $\{ A^\pm, B^\pm,C^\pm, p^\pm, r^\pm\}$, labeled by $\pm$ superscript, and chosen to depend only on the coordinates $t$ and $\varphi$. On the other hand, the two quantities $\xi^\pm$ are constants without dimensions.

The main result of this work resides in the fact that the dynamical evolution of the above geometry, according to Einstein's equations with a negative cosmological constant $R_{\mu \nu}-\frac{1}{2}R g_{\mu\nu}-\frac{1}{\ell^2}g_{\mu\nu}=0$, implies the following relations,
\begin{subequations}\label{akns}
  \begin{align}
   \pm \dot{r}^\pm+\frac{1}{\ell}\left(C'^\pm-2r^\pm A^\pm-2\xi^\pm C^\pm\right)&=0, \\
    \pm \dot{p}^\pm+\frac{1}{\ell} \left( B'^\pm+2p^\pm A^\pm+2\xi^\pm B^\pm\right)&=0, \\
    A'^\pm-p^\pm C^\pm+r^\pm B^\pm&=0.
  \end{align}
\end{subequations}
Here the dot and prime stand for the temporal and angular derivatives, respectively. Remarkably, Eqs. \eqref{akns} are two independent copies of the well-known AKNS system and therefore, two copies of the zero curvature formulation method from integrable systems \cite{Ablowitz:1973zz}. 

As shown below, this geometrization of AKNS equations is a direct consequence of a precise choice of boundary conditions for the gravitational field, i.e., the specification of the behavior of the metric field components near some surface. Although some freedom is possible when adopting boundary conditions for a gravitational theory, a reasonable choice should fulfill a number of requirements \cite{Regge:1974zd}: (i) it should render a well-defined action principle, (ii) it must be invariant under a nontrivial group of asymptotic symmetries, whose generators are finite and integrable, (iii) it must include physically interesting solutions, e.g., black holes. The guideline for the remainder of this Letter is to show the construction and consistency of boundary conditions that relate the AKNS integrable system with the dynamics of AdS$_3$ Einstein gravity.
\\

\textit{AKNS boundary conditions for the gravitational field.}
The construction of boundary conditions has much simpler pathway when carried out in the Chern-Simons formulation of AdS three-dimensional gravity \cite{Achucarro:1987vz, Witten:1988hc}. In this approach, the theory is described by the difference of two independent Chern-Simons actions with gauge group $SL(2,\mathbb{R})$ and level $k=\ell/4G$, where $G$ represents Newton's constant. The two gauge connections $\mathcal{A}^{\pm}$ are related to the dreibein $e$ and spin connection $\omega$ by $\mathcal{A}^\pm=\omega \pm e/\ell$. The first-order formulation of three-dimensional gravity is captured by the torsionless condition and the constant curvature equation (see, for example, Ref. \cite{Carlip:2004ba}). These equations, in turn, are equivalent to the zero curvature conditions $\mathcal{F}^{\pm}=0$, where $\mathcal{F}^{\pm}=d\mathcal{A}^\pm+\mathcal{A}^\pm \wedge \mathcal{A}^\pm$. The metric field can be constructed from the gauge fields as 
\begin{equation}\label{gmunu}
  g_{\mu\nu}=\frac{\ell^2}{2}\left\langle \left(\mathcal{A}_{\mu}^{+}-\mathcal{A}_{\mu}^{-}\right),\left(\mathcal{A}_{\nu}^{+}-\mathcal{A}_{\nu}^{-}\right)\right\rangle, 
\end{equation}
where $\langle\, , \, \rangle$ is the invariant bilinear form of the gauge group. The $sl\left(2,\mathbb{R}\right)$ algebra is spanned by $L_{n}$ generators, where $n\in\{-1,0,1\}$, satisfying the commutation relation
$\left[L_{n},L_{m}\right]=\left(n-m\right)L_{n+m}$. In this basis, the nonvanishing components of the invariant bilinear form are $\langle L_{1},L_{-1}\rangle=-1$ and $\langle L_{0}, L_{0}\rangle=1/2$. 
  
The boundary conditions comprise all the gauge fields of the form \cite{Coussaert:1995zp}
\begin{equation}\label{bdab}
\mathcal{A}^\pm=b^{\mp 1}( d+ a^\pm ) b^{\pm 1},
\end{equation}
where the connections $a^\pm=a^{\pm}_{\varphi}d\varphi+a^{\pm}_{t}dt$ depend only on $t$ and $\varphi$. Hence, the gauge group element $b\left(\rho\right)$ completely captures the radial dependence of the fields. Following Ref. \cite{Ablowitz:1974ry}, the angular component reads as follows,
\begin{equation}\label{aphi}
a_{\varphi}^{\pm}=\mp 2\xi^{\pm}L_{0}-p^{\pm}L_{\pm 1}+r^{\pm}L_{\mp 1}.
\end{equation}
The component along $L_0$ is chosen to be a constant without variation at the boundary.
The remaining components $p^{\pm}$ and $r^{\pm}$ are the fields carrying the boundary dynamics of the theory. In addition, the temporal component of the gauge connection is given by
\begin{equation}\label{at}
a_{t}^{\pm}=\frac{1}{\ell}(-2A^{\pm}L_{0}\pm B^{\pm}L_{\pm 1} \mp C^{\pm}L_{\mp 1}).
\end{equation}
As expected, the vanishing of the curvature two-form coincides with Eq. \eqref{akns}.

Unless stated otherwise, only the $+$ copy is treated in the following and the superscript $\pm$ is removed. Similar considerations can be applied to the $-$ copy.

A further specification of the boundary conditions is provided by choosing a precise form of the functions in $a_t$. A broad family of inequivalent choices for $A$, $B$, and $C$ with remarkable properties can be constructed recursively; these are polynomials in $\xi$ with coefficients depending on $p$, $r$, and its derivatives. In order to find them, it is useful to assume a finite expansion in powers of $\xi$,
\begin{equation}\label{rec_sol}
    A=\sum_{n=0}^{N} A_{n}\xi^{N-n},\,\, B=\sum_{n=0}^{N} B_{n}\xi^{N-n}, \,\, C=\sum_{n=0}^{N} C_{n}\xi^{N-n},
\end{equation}
where $N$ is an arbitrary positive integer. The $\xi^0$ terms in Eq. \eqref{akns} then provide the dynamical equations
\begin{equation}\label{eomN}
 \dot r = \frac{1}{\ell}\left(-C_N'+2 r A_N\right),\quad  \dot p = \frac{1}{\ell}\left(-B_N'-2 p A_N\right).
\end{equation}
The remaining terms imply the following recursion relations for the coefficients in the expansion
\begin{subequations}\label{recurrence}
  \begin{align}
    A_{n}'&=pC_{n}-rB_{n}, \label{recurrenceA}\\
    B_{n+1}&=-\frac{1}{2}B_{n}'-p A_{n},  \\
    C_{n+1}&=\frac{1}{2}C_{n}'-r A_{n},
  \end{align}
\end{subequations}
along with $B_0=C_0=0$. To find an explicit solution to the recursion relations, it is useful to take into account of the conserved quantities of the AKNS system. Indeed, equations \eqref{akns} imply the existence of an infinite set of conserved charges $H_n$, with $n\in \mathbb{N}$, which can be obtained by a means of a recursive manipulation of the AKNS equations \cite{Ablowitz:1974ry}. The first quantities read
\begin{equation}
  \mathcal{H}_1=0,\quad\mathcal{H}_2=-pr,\quad \mathcal{H}_3=\frac{1}{4}(p'r-pr')\quad...
\end{equation}
where $H_n=\int \mathcal{H}_n \,d\varphi$. The trivial quantity $\mathcal{H}_1$ is appended for notation reasons. The recursion relations \eqref{recurrence} can then be explicitly solved for each coefficient \cite{trace}, yielding $B_0=C_0=0$, $A_0=1$ and
\begin{equation} \label{rec_solution2}
    A_n=\frac{n-1}{2}\mathcal{H}_n,\quad B_n=\mathcal{R}_{n+1},\quad C_n=\mathcal{P}_{n+1},
  \end{equation}
for $n \geq 1$. All integration constants are fixed to zero, except $A_{0}=1$. This choice does not alter the forthcoming analysis. The quantities $\mathcal{R}_{n}$ and $\mathcal{P}_{n}$ correspond to variational derivatives of the conserved charges $H_n$   
\begin{equation}
\mathcal{R}_n \equiv \frac{\delta H_n}{\delta r} , \quad \mathcal{P}_n \equiv \frac{\delta H_n}{\delta p}.  
\end{equation}

From the above discussion it is clear that different values of the positive integer $N$ give rise to distinct dynamics. Hence, inequivalent choices of the boundary conditions are labeled by the integer $N$. Several well-known integrable equations arise as particular cases of the above construction: Korteweg-de Vries ($N=3$, $r=1$), modified Korteweg-de Vries ($N=3$, $r=p$), (Wick rotated) nonlinear Schr\"{o}dinger ($N=2$), chiral boson ($N=1$), among others. The Sine-Gordon equation is also included in this framework, however, negative powers of $\xi$ must be included in the expansion in order to make it apparent. This case and its connection to gravity will be addressed elsewhere. 

As a final comment for this section, it should be noted that a bi-Hamiltonian structure for Eq. \eqref{eomN} can be revealed by casting the equations conveniently as    
\begin{equation}
  \left(
    \begin{aligned}
   \dot r\\  \dot p
    \end{aligned}
\right) = \mathcal{D}_1 \left(
    \begin{aligned}
      \mathcal{R}_{N+1}\\ \mathcal{P}_{N+1}
    \end{aligned}
\right).
\end{equation}
The first Hamiltonian operator $\mathcal{D}_1$ can be read from Eq. \eqref{recurrence}, yielding
\begin{equation}
  \mathcal{D}_1=\frac{1}{\ell}\left(
    \begin{matrix}
      -2r \d_\varphi^{-1}(r\cdot \; ) & -\d_\varphi +2r\d_\varphi^{-1}(p\cdot \; ) \\
      -\d_\varphi +2p\d_\varphi^{-1}(r\cdot \; ) & -2p \d_\varphi^{-1}(p\cdot \; )
    \end{matrix}
\right).
\end{equation}
Alternatively, by virtue of the recurrence relation \eqref{recurrence}, the dynamical equations can also be written as
\begin{equation}
  \left(
    \begin{aligned}
    \dot r\\ \dot p
    \end{aligned}
\right) = \mathcal{D}_2 \left(
    \begin{aligned}
      \mathcal{R}_{N+2}\\ \mathcal{P}_{N+2}
    \end{aligned}
\right),
\end{equation}
where the second Hamiltonian operator is given by
\begin{equation}
  \mathcal{D}_2=\frac{1}{\ell}\left(
    \begin{matrix}
      0 & -2\\
      2 & 0
    \end{matrix}
\right).
\end{equation}
The operators $\mathcal{D}_1$ and $\mathcal{D}_2$ are compatible, in the sense that the combination $  \mathcal{D}_1+\mathcal{D}_2$ is also a Hamiltonian operator \cite{Olver}. This observation is intimately related to the recurrence relation \eqref{recurrence}, which can then be expressed as
\begin{equation}\label{bihamrec}
  \left(
    \begin{aligned}
      \mathcal{R}_{n+1}\\ \mathcal{P}_{n+1}
    \end{aligned}
\right) = \mathcal{D}_2^{-1}\mathcal{D}_1 \left(
    \begin{aligned}
      \mathcal{R}_{n}\\ \mathcal{P}_{n}
    \end{aligned}
\right).
\end{equation}
\\

\textit{Consistency of the boundary conditions.}
The above construction provides a complete framework to address the question whether Eqs. \eqref{bdab},\eqref{aphi}, and \eqref{at} define an adequate set of boundary conditions.

Concerning the construction of a well-defined action principle, the Chern-Simons action should be supplemented with a boundary term $\mathcal{B}$ such that the variation of the action vanishes on shell. The variation of such boundary term reads   
 \begin{equation}
  \delta \mathcal{B} =-\frac{k}{2 \pi} \int dt d\varphi \left<a_t, \delta a_\varphi   \right>.
\end{equation}
By virtue of Eq. \eqref{rec_solution2}, this expression readily integrates to
\begin{equation}
  \mathcal{B} =\frac{k}{2 \pi} \int \frac{dt}{\ell} \sum_{n=0}^N \xi^{N-n}H_{n+1},
\end{equation}
which then renders the action differentiable, as needed.

Regarding asymptotic symmetries, they correspond to the family of infinitesimal gauge transformations 
\begin{equation} \label{gauge-trans}
\delta a = d \Lambda + \left[ a,\Lambda \right],
\end{equation}
which respect the form of the boundary conditions \eqref{aphi} and \eqref{at}. In order to find them, consider a general gauge parameter $ \Lambda=-2\alpha L_{0}+\beta L_{1}-\gamma L_{- 1} $. The angular component of the transformation \eqref{gauge-trans} yields equations analogous to Eq. \eqref{akns}. As a result, the functions $\alpha$, $\beta$, and $\gamma$ are
\begin{subequations} \label{gauge-param}
\begin{align}
    &\alpha=\sum_{n=0}^{M} \frac{(n-1)}{2}\mathcal{H}_n\xi^{M-n},\,\,  \\
    &\beta=\sum_{n=0}^{M}\mathcal{R}_{n+1}\xi^{M-n}, \,\,\\
    & \gamma=\sum_{n=0}^{M} \mathcal{P}_{n+1}\xi^{M-n},
\end{align}
\end{subequations}
where $M$ is a positive integer (not necessarily equal to N). Here, $M$ labels an infinite family of permissible gauge transformations. The infinitesimal transformation of the fields $r$ and $p$ are then given by 
\begin{equation}\label{eomM}
  \delta r = -\gamma_M'+2 r \alpha_M,\quad \delta p = -\beta_M'-2 p \alpha_M.
\end{equation}

From the Hamiltonian point of view, the gauge transformation \eqref{gauge-trans} is generated by the boundary term $Q[\Lambda]$ that must be supplemented to the first class constraint in order to yield it differentiable \cite{Banados:1994tn}. Its variation is given by 
\begin{equation}\label{deltaQ}
\delta Q[\Lambda]=\frac{k}{2\pi} \int d\varphi \left( \beta \delta r+\gamma \delta p\right),
\end{equation}
which, by virtue of Eq. \eqref{gauge-param}, can be integrated to
\begin{equation}\label{Q}
 Q[\Lambda] =\frac{k}{2 \pi} \sum_{n=0}^M \xi^{M-n}H_{n+1}.
\end{equation}
The generators of this family are in involution, i.e., span an Abelian algebra $\{Q[\Lambda],Q[\bar \Lambda]\}=0$, where the brackets stand for the canonical Poisson bracket\footnote{The proof stems from the fact that the charges canonically generate the transformations $\{Q[\Lambda],Q[\bar \Lambda]\}=\bar \delta Q[\Lambda]$ \cite{Brown:1986ed}. Considering \eqref{deltaQ}, the right hand side of the latter expression can then be written as
  $$
\{Q[\Lambda],Q[\bar \Lambda]\}=\frac{k\ell}{2\pi}  \sum_{n=0}^M \xi^{M-n}\int d\varphi
\begin{pmatrix}
  \mathcal{R}_{n+1} & \mathcal{P}_{n+1}
\end{pmatrix}
\mathcal{D}_1
\begin{pmatrix}
  \mathcal{R}_{\bar M +1} \\ \mathcal{P}_{\bar M +1}
\end{pmatrix}.
   $$
The term inside the integral vanishes by virtue of the recurrence relation \eqref{bihamrec}.}.

It is clear by construction that $a_t$ belongs to the above family of permissible gauge parameters $a_t=\frac{1}{\ell}\Lambda|_{M=N}$. Consequently, the time evolution of the system also respects the boundary conditions, as it should. In addition, regarding the effect of the asymptotic symmetry transformation $\Lambda$ acting on $a_t$, the temporal component of Eq. \eqref{gauge-trans} reduces to combinations of the equations of motion \eqref{eomN} and the infinitesimal transformations of the fields \eqref{eomM}. Thus, it does not imply any further condition on the gauge parameter\footnote{A more general statement can be proved as follows. Consider two asymptotic symmetry transformations with parameter $\Lambda$ and $\bar \Lambda$. It is straightforward to show that its commutation yields a third transformation
  $
(\delta \bar \delta - \bar \delta \delta)a_\varphi= \bar {\bar \delta} a_\varphi,
$
with parameter $\bar{ \bar \Lambda}=\delta \bar \Lambda-\bar \delta \Lambda+\left[\Lambda,\bar \Lambda  \right]$. However, by virtue of Jacobi identity and the involution of the canonical generators \eqref{Q} under Poisson brackets, the commutation of two such transformations vanishes, thus implying $\delta \bar \Lambda-\bar \delta \Lambda+\left[\Lambda,\bar \Lambda  \right]=0$. The temporal component of \eqref{gauge-trans} is obtained from the latter expression when $\bar M = N$.  
}.

To finalize this section, let us briefly discuss some properties of the spacetime geometry defined by Eqs. \eqref{lapse}, \eqref{shift}, and \eqref{gij}. The spacetime metric can be constructed by means of Eq. \eqref{gmunu}, considering a radial group element chosen as $b^\pm(\rho)= \exp \left[ \pm \log \left( \frac{\rho}{\ell}\right)  L_0 \right] $. Note that the boundary dynamics arises from the asymptotic behavior of the lapse function and shift vector \cite{Henneaux:2013dra}, which depend on the dynamical functions and consequently induce a nontrivial surface evolution at the boundary. This property has been used previously in Refs. \cite{Perez:2016vqo} and \cite{Ojeda:2019xih} to connect the dynamics of AdS$_3$ general relativity with the KdV and Gardner integrable hierarchies, respectively.    

The parameters $\xi^\pm$ are absent from the field equations \eqref{eomN}. However, for a given solution of Eq. \eqref{eomN}, different values for the parameters corresponds to nonequivalent geometries. Indeed, classical solutions in three-dimensional gravity can be classified by the trace of the holonomy of the gauge connection along the angular cycle $M^{\pm}=\mbox{Tr}\left(\mathcal{P}\;\mbox{exp}\oint A^{\pm}_\varphi d\varphi \right)$, which is a gauge invariant quantity. It characterizes the three conjugacy classes of $SL(2,\mathbb{R})$, describing different kinds of spacetimes \cite{Martinec:1998wm}. In the present case, it yields
\begin{equation}
M^\pm =2\cosh\left(2 \pi \sqrt{(\xi^{\pm})^2+p^{\pm}_{0}r^{\pm}_{0}}\right),
\end{equation}
where, $p^{\pm}_{0}$ and $r^{\pm}_{0}$ are the zero modes of the Fourier expansions $p^{\pm}=\sum_n p_n^\pm \mbox{exp}(i n \varphi)$ and $r^{\pm}=\sum_n r_n^\pm \mbox{exp}(i n \varphi)$. If $M^{\pm}<2$, the configuration represents the elliptic conjugacy class and corresponds to classical particle sources, inducing conical singularities. The case $M^{\pm}>2$ typifies hyperbolic elements of $SL(2,\mathbb{R})$ that characterize black hole solutions. The last possible scenario, $M^{\pm}=2$, leads to parabolic conjugacy classes and describe extremal black holes.

Remarkably, the three above configurations are attainable. If $p_0^\pm$ and $r_0^\pm$ are both positive or both negative, the solutions generically represent non stationary black holes. Conical singularities and extremal black holes are obtained if $p_0^\pm$ and $r_0^\pm$ have opposite signs, while selecting a suitable value for $\xi^\pm$, which as mentioned before, is not fixed by the dynamical equations.

Once a solution of the equations of motion \eqref{eomN} is found, the geometry described by Eqs. \eqref{lapse}, \eqref{shift}, and \eqref{gij}, renders a solution of three-dimensional Einstein’s equations. Hence, it corresponds to a locally AdS spacetime. The global properties described by zero modes can be constructed by means of identifications of global AdS \cite{Banados:1992gq}. Higher order modes, characterized by AKNS charges, correspond to large gauge transformations \eqref{eomM}. Furthermore, there are nonpermissible gauge transformations that do not preserve the boundary conditions \eqref{aphi} and \eqref{at}, such as locally recasting the metric into different forms, for example, into a conformally flat spacetime or near horizon geometries \cite{Afshar:2016wfy}. Their action map spacetimes with different asymptotic behavior leading to distinct conserved charges and symmetry algebras.  One example of this issue was found in Ref. \cite{Afshar:2016wfy}, where the relationship between Brown-Henneaux and the soft hairy black holes boundary conditions was established. Analogously, integrable systems can also be related through a gauge transformation, like the nonlinear Schr\"odinger and Landau-Lifshitz equation \cite{gaugeequi}. The connection between these gauge-related integrable systems and the nonpermissible gauge transformation in the gravitational formalism can be studied from both angles.

\textit{Final remarks.}
The family of boundary conditions constructed here encompasses some examples found in the literature. The well-known Brown-Henneaux case \cite{BrownHenneaux} can be recovered if $N^\pm=1$, $r^\pm=1$, and then setting $\xi^\pm=0$. Additionally, the family of KdV boundary conditions found in Ref. \cite{Perez:2016vqo} is recovered for $r^\pm=1$, odd values of $N^\pm$ and vanishing $\xi^\pm$. The relation between this hierarchy and black hole properties is studied in Refs. \cite{Grumiller:2017jft,Erices:2019onl,Dymarsky:2020tjh}. A detailed discussion of how this work relates to several other boundary conditions for AdS$_3$ gravity \cite{Compere:2013bya,Ojeda:2019xih,Grumiller:2016pqb,Troessaert:2013fma} will be given elsewhere.

Regarding the physical significance of the infinite conserved charges, it is worth mentioning that if a gravitational configuration describing a black hole is endowed with nonzero AKNS charges, these can be considered ``soft hair'' \cite{Hawking:2016msc} in the following sense. The conserved quantities commute with the total gravitational Hamiltonian, which is given by $Q^+[a^+_t]+Q^-[a^-_t]$. Hence, the gauge transformations generated by $Q^\pm[\Lambda^\pm]$ map a black hole configuration into a physically inequivalent one, without changing the energy of the system.

Several questions arise concerning the gravitational counterpart of well-established methods and results in the literature. For example, what is the meaning of the different generating solution schemes, e.g., Darboux transformations or Hirota method, in the gravity side?  How are these results connected with the self-dual Yang Mills description of integrable systems? Is there a gravity analog for those integrable systems related by gauge transformations? 

Further generalizations of this work can be pursued. To mention a few, the case of vanishing cosmological constant should be related to the integrable hierarchy constructed out of non-semi-simple algebras \cite{Shen2017} that include the Poincar\'e case. Similarly, three-dimensional higher spin gravity on AdS$_3$ \cite{Blencowe:1988gj, Bergshoeff:1989ns} should be related to the integrable dynamics of $sl(N,\mathbb{R})$ generalizations of the AKNS system. See Ref. \cite{Ojeda:2020bgz} for a recent example.
\\

\textit{Acknowledgments.}
The authors thank Hern\'an A. Gonz\'alez and Julio Oliva for useful discussions and comments. This research has been partially supported by FONDECYT Grants 1181628,  11190730, 1171475, 1211356 and the grant ANID Beca Doctorado Nacional 21182110. F.C. would like to thank the Departamento de F\'isica at the Universidad de Santiago de Chile for warm hospitality.



\end{document}